\documentclass{sigchi}

\usepackage{balance}       
\usepackage{graphics}      
\usepackage[T1]{fontenc}   
\usepackage{txfonts}
\usepackage{mathptmx}
\usepackage[pdflang={en-US},pdftex]{hyperref}
\usepackage{color}
\usepackage{booktabs}
\usepackage{textcomp}
\usepackage{subcaption}
\usepackage{epstopdf}
\usepackage{microtype}        
\usepackage{ccicons}          

\usepackage{todonotes}

\def\plaintitle{Can We Predict the Scenic Beauty of Locations from Geo-tagged Flickr Images?}

\def\emptyauthor{}
\def\plainkeywords{Data Mining, Machine Learning, Flickr, Social Metadata.}

\makeatletter
\def\url@leostyle{%
  \@ifundefined{selectfont}{
    \def\UrlFont{\sf}
  }{
    \def\UrlFont{\small\bf\ttfamily}
  }}
\def\@copyrightspace{\relax}
\makeatother
\def\pprw{8.5in}
\def\pprh{11in}

\definecolor{linkColor}{RGB}{6,125,233}
\hypersetup{%
  pdftitle={\plaintitle},
  pdfauthor={\emptyauthor},
  pdfkeywords={\plainkeywords},
  pdfdisplaydoctitle=true, 
  bookmarksnumbered,
  pdfstartview={FitH},
  colorlinks,
  citecolor=black,
  filecolor=black,
  linkcolor=black,
  urlcolor=linkColor,
  breaklinks=true,
  hypertexnames=false
}

\begin{document}

\title{\plaintitle}

\numberofauthors{2}
\author{%
  \alignauthor{Ch. Md. Rakin Haider\\
    \affaddr{Bangladesh University of Engineering and Technology,}
    \affaddr{Dhaka, Bangladesh}\\
    \email{rakinhaider@cse.buet.ac.bd}}\\
  \alignauthor{Mohammed Eunus Ali\\
  	\affaddr{Bangladesh University of Engineering and Technology,}
    \affaddr{Dhaka, Bangladesh}\\
    \email{eunus@cse.buet.ac.bd}}\\
}

\maketitle

\begin{abstract}
In this work, we propose a novel technique to determine the aesthetic score of a location from social metadata of Flickr photos. In particular, we built machine learning classifiers to predict the class of a location where each class corresponds to a set of locations having equal aesthetic rating. These models are trained on two empirically build datasets containing locations in two different cities (Rome and Paris) where aesthetic ratings of locations were gathered from \textit{TripAdvisor.com}. In this work we exploit the idea that in a location with higher aesthetic rating, it is more likely for an user to capture a photo and other users are more likely to interact with that photo. Our models achieved as high as 79.48\% accuracy (78.60\% precision and 79.27\% recall) on Rome dataset and 73.78\% accuracy(75.62\% precision and 78.07\% recall) on Paris dataset. The proposed technique can facilitate urban planning, tour planning and recommending aesthetically pleasing paths.
\end{abstract}

\category{}{Information systems}{Data mining}
\category{}{Information systems}{Social networks}
\category{}{Computing methodologies}{Supervised learning}


\section{Introduction}

Aesthetic ratings of locations play a vital role in  urban planning and tour planning. Satisfying a certain time and budget constraints, each tourist wishes to visit most attractive places of a city.  Such interest have resulted in the increasing popularity of websites like TripAdvisor, Expedia and Travelocity. They help a tourist to choose the best tourist spots in a city. One of the most popular trip recommendation site, TripAdvisor, generates the location ratings based on crowd-sourced reviews and ratings for each location \cite{TripAdvisorPopularity}. In this paper, we propose a novel technique to predict TripAdvisor location ratings from the social metadata of photos available in a major social content sharing website, Flickr. 

The last decade has witnessed an unprecedented rise in the usage of GPS technology and the availability of GPS-enabled camera phones have increased creation of geo-tagged photos. Popular content sharing sites such as Flickr, Instagram and Twitter have already accumulated a large number of user generated photos. In Flickr alone there are about 8 billion existing photos and it is reported that Flickr receives 3.5 million uploads every day. These photos contain metadata such as gps-location, time of the photo taken, number of users viewed and number of favorites. The large amount of geo-tagged photos not only gives us plenty of information about a specific geo-location\cite{JiGZYT11,SpyrouM16,HollensteinP10} but also contains patterns of human behavior, photo trails\cite{ArasePhotoTripMining,shafique2016recommending}, and even transit time \cite{PopescuG09} between areas of a city. The increased availability of such geographic information in the form of multimedia content such as images have given rise to interesting technologies such as recommendation system, point-of-interest mining, tour planning system etc. Apart from trip planning and route recommendations, research works have also been conducted to find out quantity of ecological phenomena like snow-fall and vegetation density \cite{ZhangKCL12}, predicting user's home location and gender \cite{PopescuG10}, detecting events that took place \cite{RattenburyGN07} etc. each of which used Flickr geo-tagged photos and metadata. In most of these studies geo-tags of photos, temporal metadata, content-based automatically generated tags as well as user generated tags were considered for the task of analysis. However, Flickr contains another set of metadata which are generated from users interaction with the photos and with other users. A few such social metadata are number of views, number of favorites, comments etc. These metadata are mostly overlooked in recent studies.

In this work, we propose a method to predict the aesthetic score of a location from Flickr social metadata. To the best of our knowledge this is the first attempt to predict TripAdvisor location ratings from Flickr social metadata. We exploited the idea that photo capturing patterns in scenic locations should be significantly different than that of lesser ones. We have modeled the problem as a multi-class classification problem were each location is a member of one of the equally rated classes. In TripAdvisor, the ratings are provided on a scale of 0 to 5 with an interval of 0.5. We considered 6 classes with aesthetic rating 2.5, 3, 3.5, 4, 4.5 and 5 respectively and discarded locations with lesser aesthetic ratings. In Paris dataset we have considered an additional class with aesthetic rating 2.0. These classification models can be used to provide suggestions to tourists, artist and urban resource management organizations.

The leading trip recommendation system, TripAdvisor, provides location ratings based on the 500 million reviews and opinions about 7 million attractions, accumulated from 300 million of its users. TripAdvisor considers three factors i.e. quality, recency and quantity of the reviews while computing the ratings. A major drawback of this method of aesthetic rating calculation is the dependancy on crowd-sourced reviews. It is likely that a tourist will prefer to post his photos on social media rather than rating each location he visited in order to help generating aesthetic ratings. The goal of this work is to reduce the dependancy on crowd-sourced ratings. We used aesthetic ratings calculated from TripAdvisor data as ground truth for our work and tried to predict them from the data avaiable in other forms. In this work, we have gathered TripAdvisor ratings of 850 locations in Rome, Italy and 650 locations in Paris, France. For each of the locations we have collected social metadata of photos which are taken within a radius of 100m of that geo-location. We have gathered social metadata of 6 million photos in Rome and 4 million photos in Paris using Flick API. From these metadata we have generated some relevant features for classification. Since significant imbalance was observed in the dataset of both cities, we have applied state-of-art technique, namely SMOTE \cite{SMOTE} to balance the classes out. We have trained multiple decision tree variants as our classifier i.e. J48, Random Forrest(RF) and REPTree. Finally, application of ensemble technique further contributed to increasing accuracy. Among the trained models, Random Forrest performed best when combined with bagging ensemble method and achieved 79.48\% accuracy on Rome dataset. On the other hand, the best performance(73.78\% accuracy) is found when Random Forrest classifier is combined with boosting ensemble technique in Paris dataset.

The major contributions of this work are summarized below.
\begin{itemize}
\item We propose a novel approach to predict location ratings from photo capturing, sharing and interaction patterns.
\item We build two datasets empirically to facilitate TripAdvisor location rating prediction for the locations of Rome, Italy and Paris, France. 
\item We train decision tree based classifiers i.e. J48, random forrest, REPTree, whose performance was improved with the help of bagging and boosting ensemble technique. We also handle issues regarding imbalanced dataset by applying SMOTE \cite{SMOTE} oversampling technique .  
\item We test our models on two real world datasets related to two popular tourist destinations, Rome, Italy and Paris, France. Our models demonstrated 79.48\% and 73.78\% accuracy respectively. 
\end{itemize}

\section{Methodology}

\subsection{Dataset Generation}
There are multiple available datasets related to Flickr that can be considered while choosing the dataset. The first one of them is the \textit{yfcc100m} dataset publised by Yahoo Webscope \cite{yfcc100m}. This dataset contains a collection of photos and videos, which are compiled from the data available on Yahoo! Flickr. The dataset is divided into three parts. The main part of the dataset contains information about 100 million flickr photos. It contains photo or video identifier, photo/video hash, user information, date in which the photo was taken, upload date, title of the photo, description, user tags (comma-separated), location information, specifications of the device by which the photo or video was captured and URLs of the photo or video. \textit{yfcc100m} also includes machine tags and human readable place information for every photo. An alternative of \textit{yfcc100m} dataset is the Multimedia Commons Repository\textit{(MMC)} \cite{thomee2016yfcc100m}. The differences between \textit{yfcc100m} and \textit{MMC} are the supplemental material to \textit{yfcc100m} that the \textit{MMC} offers. \textit{MMC} offers audio, visual and motion features such as LIRE, GIST, SWIFT that are often used by multimedia researchers. There are several other sources such as MIRFLICKR, Flickr API etc., through which one can access large number of flickr photos and their related attributes. 

In order to perform the desired task of classification we need a dataset that provides social metadata such as number of views, number of favorites, number of comments etc. of Flickr photos as well as ground truth with respect to the aesthetic scores of geo-locations. From the discussion about \textit{yfcc100m} and \textit{MMC}, we can observe that none of them include social metadata of the photos. Moreover, none of these datasets include any ground truth about a place being aesthetically desirable. To facilitate our classification task we decided to build a new dataset using Flickr social metadata and aesthetic rating of geo-locations from TripAdvisor. 

In order to appropriately model photo distribution around a city or country we tried to gather location names and ratings from \textit{TripAdvisor.com}. Since trip advisor does not provide API for research purposes, we used HTML parser to get data from Trip Advisor. From Trip Advisor we have retrieved rating, and number of reviews for around 1200 attractions in Rome, Italy and 1200 attractions in Paris, France. After removing duplicates and attractions such as tours, restaurants and hotels we were left with 850 and 650 locations for consideration respectively. Despite the provided ratings being on a scale of 0-5 with an interval of 0.5, we observed from the data, that most of the ratings associated with top attractions in Rome lies in the range 2.5-5 and within 2-5 in Paris. For each location we retrieved latitude and longitude using \textit{Google Places API}. Then for each geo-location we fetched metadata of all the images that lie within a circle of 100m radius  surrounding that particular location from Flick API. Thus we set up two datasets where each location is labeled as one of 7 classes i.e. classes with aesthetic scores \{2,2.5,3,3.5,4,4.5,5\}. Let the classes be named as $C_i$ where $i$ is the corresponding aesthetic score. Among the locations in Rome there are points-of-interests(197), museums(130), church and cathedrals(78), historic sites(74), castles(25), gardens(23), parks(10), neighborhoods(13) etc. Similarly, the types of locations considered in Paris are points-of-interests(115), museums(68), neighborhoods(39), church, cathedrals and historic sites (28) and so on.

\subsection{Feature Extraction}
The social metadata related to each photo that are directly available from Flickr are the number of times the photo is viewed, the number of people who have added the photo as favorite, and the number of comments on the photo. We have extracted some aggregate features using these metadata for each location. The major intuitions behind our features are,
\begin{itemize}
\item A place with more aesthetic beauty encourages more users to take photos and upload them, resulting in a higher density of photos at that place.
\item An aesthetically beautiful place is more likely to draw tourists and the number of distinct users uploading photos at a place should be higher.
\item Visitors usually searches for and views photos captured at beautiful locations.
\item The more beautiful a place is, it is more likely that people will capture better photos which results in higher number of people adding them to favorites.
\item Higher number of views should lead to higher number of favorites and comments.
\item The quality of photos, taken at an aesthetic and popular location, usually depends on the surrounding scenario rather than the photographer's skills.
\end{itemize}

\begin{figure}[!ht]
	\centering
	    \begin{subfigure}[b]{0.15\textwidth}
        \includegraphics[width=\textwidth]{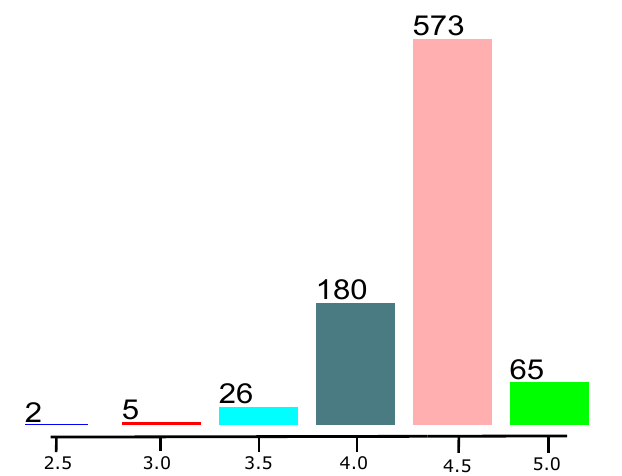}
        \caption{Frequency histogram locations per each aesthetic rating}
        \label{fig:statusbar}
    \end{subfigure}
	\begin{subfigure}[b]{0.15\textwidth}
        \includegraphics[width=\textwidth]{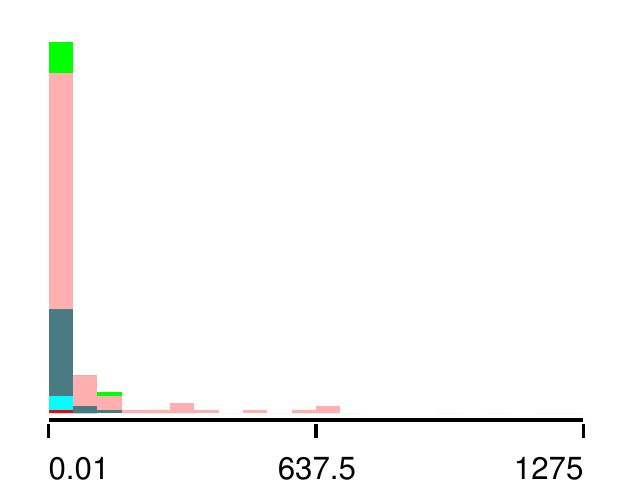}
        \caption{Frequency histogram of feature "photo desity"}
    \end{subfigure}
    \begin{subfigure}[b]{0.15\textwidth}
        \includegraphics[width=\textwidth]{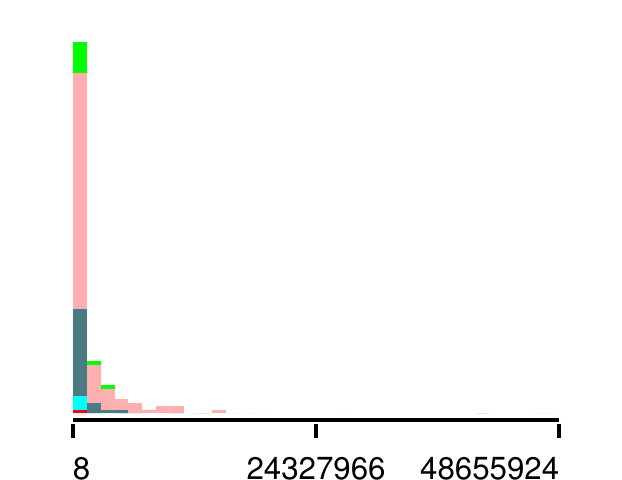}
        \caption{Frequency histogram of feature "total number of views"}
    \end{subfigure} \\

    \begin{subfigure}[b]{0.15\textwidth}
        \includegraphics[width=\textwidth]{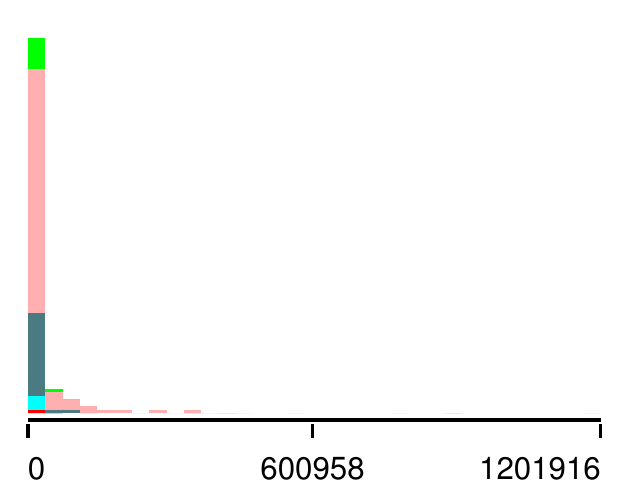}
        \caption{Frequency histogram of feature "total number of favourites"}
    \end{subfigure}
    \begin{subfigure}[b]{0.15\textwidth}
        \includegraphics[width=\textwidth]{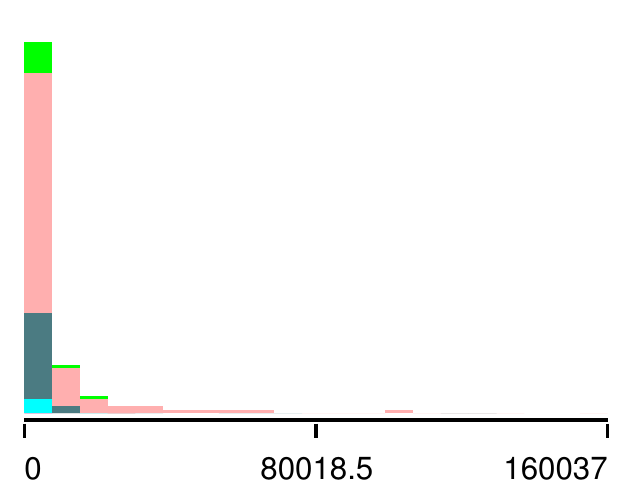}
        \caption{Frequency histogram of feature "total number of comments"}
    \end{subfigure}
    \begin{subfigure}[b]{0.15\textwidth}
        \includegraphics[width=\textwidth]{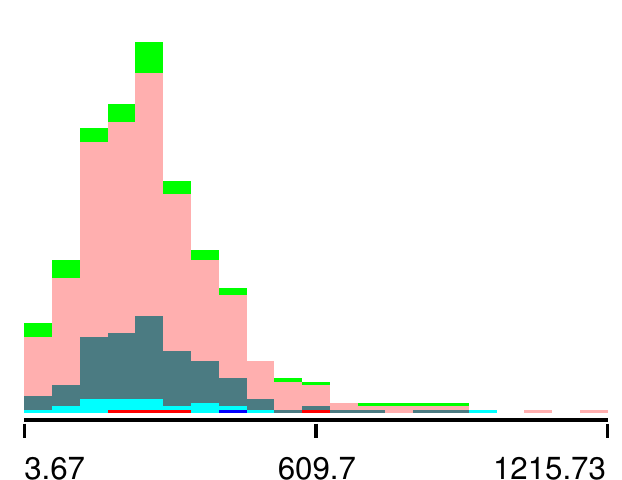}
        \caption{Frequency histogram of feature "average views per photo"}
    \end{subfigure}\\
    
	\begin{subfigure}[b]{0.15\textwidth}
        \includegraphics[width=\textwidth]{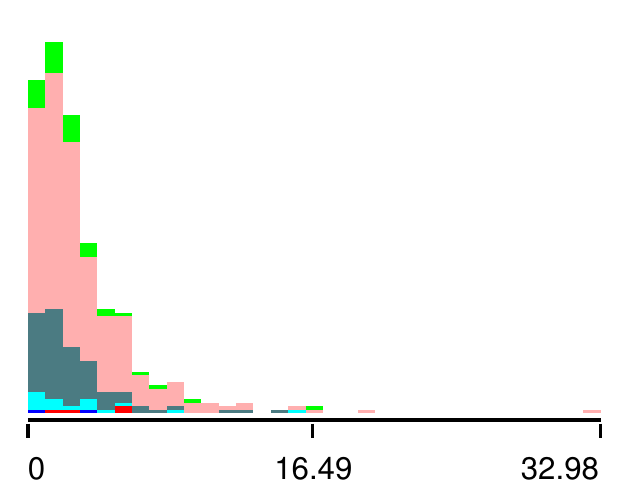}
        \caption{Frequency histogram of feature "average number of favourites per photo"}
    \end{subfigure}
    \begin{subfigure}[b]{0.15\textwidth}
        \includegraphics[width=\textwidth]{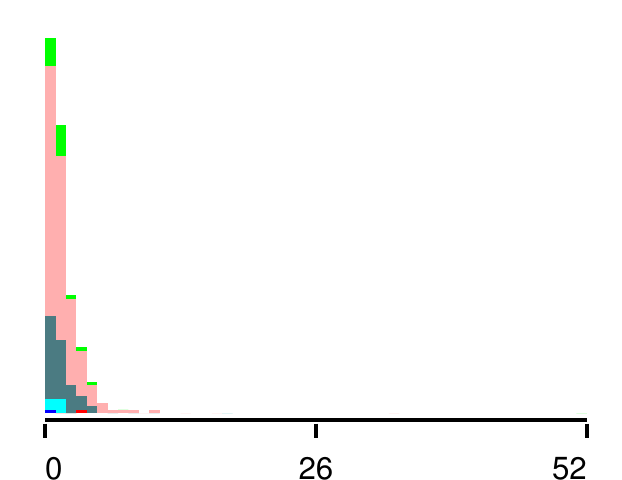}
        \caption{Frequency histogram of feature "average number of comments per photo"}
    \end{subfigure}
    \begin{subfigure}[b]{0.15\textwidth}
        \includegraphics[width=\textwidth]{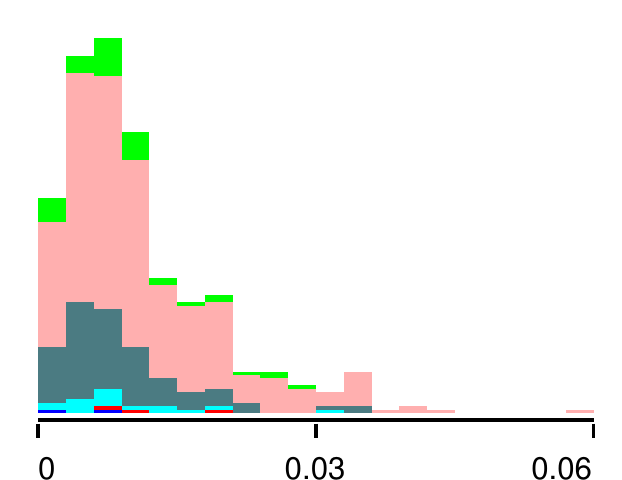}
        \caption{Frequency histogram of feature "favourite to view ratio"}
    \end{subfigure}\\
    
    \begin{subfigure}[b]{0.15\textwidth}
        \includegraphics[width=\textwidth]{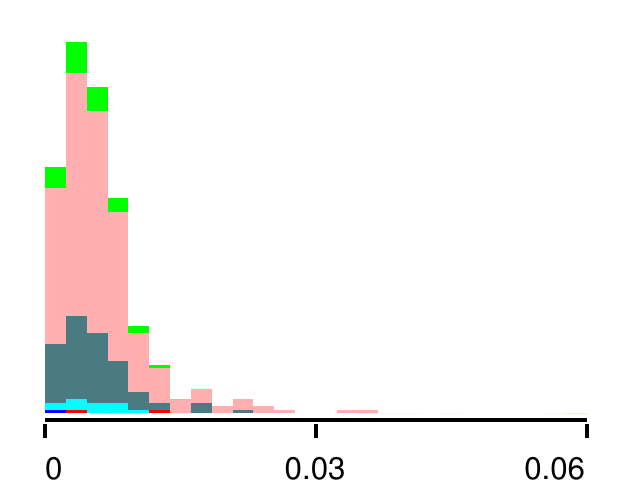}
        \caption{Frequency histogram of feature "comment to view ratio"}
    \end{subfigure}  
    \begin{subfigure}[b]{0.15\textwidth}
        \includegraphics[width=\textwidth]{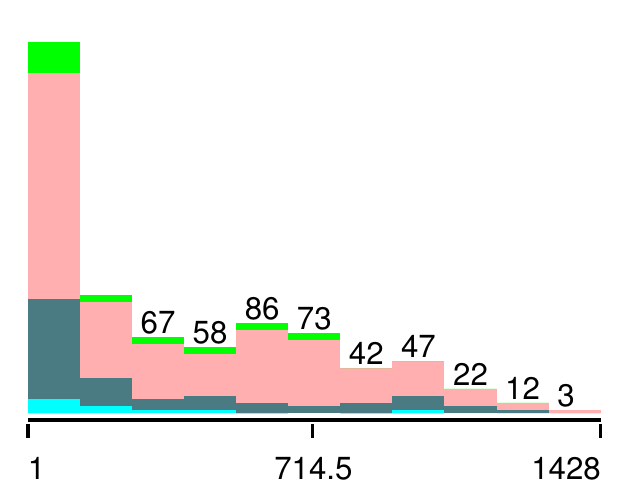}
        \caption{Frequency histogram of feature "number of distinct users"}
    \end{subfigure}
    \begin{subfigure}[b]{0.15\textwidth}
        \includegraphics[width=\textwidth]{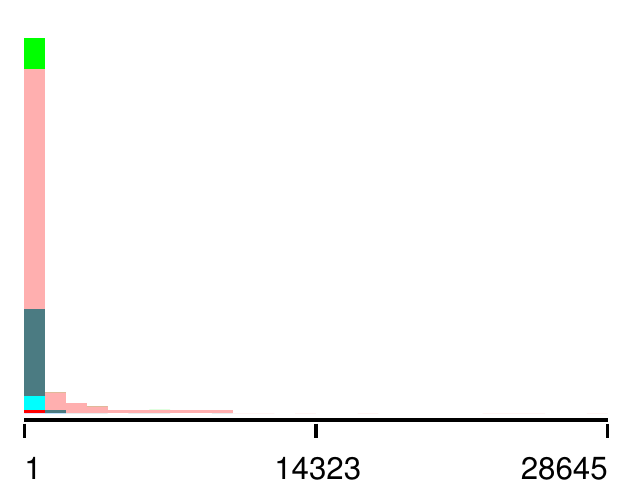}
        \caption{Frequency histogram of feature "maximum number of photo per user in a location"}
    \end{subfigure}
    \caption{Visual Representation of distribution of the features.}
    \label{fig:visualization}
\end{figure}

Keeping these points in mind we have generated 11 features for every location. The generated features for each location are photo density, total number of views, total number of favorites, total number of comments, average views per photo, average favorite count, average number of comments per photo, ratio of number of favorites to number of views, ratio of number of comments to number of views, distinct user count per location, and finally maximum number of photo per user.
Figure \ref{fig:visualization} shows the frequency histograms of each feature generated from Rome dataset. The number of instances of a class per histogram bar is represented with appropriate colored portions. Additionally, Figure \ref{fig:statusbar} shows the number of locations for each aesthetic score and their associated color used in the histograms. 
\subsection{Classification}
We have trained several classifiers to be able to classify the places as a member of one of the six(or seven in case of Paris dataset) equally rated classes. We evaluated each of their accuracy, precision and recalls to distinguish the best classifier. As our classifiers, we have used J48, REPTree and Random Forrest(RF) decision tree algorithm. To boost the accuracy of classifiers we have also applied Bagging and Boosting technique on the decision tree classifiers.

\section{Experimental Results}
\subsection{Classification} \label{sec:results}
We have considered aesthetic scores as nominal values where a location can have one of 6 aesthetics scores in Rome dataset, starting from 2.5 to 5 with an interval of 0.5. Similarly, a location can have one of 7 possible aesthetic score in Paris dataset, from 2.0 to 5 with 0.5 interval. To solve the classification task we have trained three classifier. The classifiers are different variations of decision tree classifier. They are J48, REPTree and Random Forrest. We have applied 10 fold cross-validation technique to validate the accuracy of classifiers. Table \ref{tab:decisionTreeResultsRome} and \ref{tab:decisionTreeResultsParis} show the accuracy, precision and recall of each classifier. According to Sokolova et al.\cite{PerformanceMeasureClassifier} the precision and recall of multi-class classifier are the average precision and average recall of all the individual classes.    
\begin{center}
	\begin{table}[!h]
		\caption{Classifier accuracy, precision and recall of decision tree classifiers on dataset with 851 locations of Rome.}
		\label{tab:decisionTreeResultsRome}
		\centering
		\begin{tabular}{||c | c | c | c||}
			\hline
			Classifier & Accuracy & Precision & Recall\\
			\hline\hline
			J48 &	64.39 &	Undefined &	17.60 \\
			\hline
			REPTree &	66.51 &	Undefined &	16.78 \\
			\hline
			RF &	60.28 &	Undefined &	19.57 \\
			\hline
		\end{tabular}
	\end{table}
\end{center}
\vspace{-2em}
\begin{center}
	\begin{table}[!h]
		\caption{Classifier accuracy, precision and recall of decision tree classifiers on dataset with 650 locations of Paris.}
		\label{tab:decisionTreeResultsParis}
		\centering
		\begin{tabular}{||c | c | c | c||}
			\hline
			Classifier & Accuracy & Precision & Recall\\
			\hline\hline
			J48 &	36.09 &	Undefined &	18.65 \\
			\hline
			REPTree &	41.44 &	Undefined &	16.04 \\
			\hline
			RF &	42.81 &	Undefined &	19.46 \\
			\hline
		\end{tabular}
	\end{table}
\end{center}
\vspace{-2.5em}
The results in Table \ref{tab:decisionTreeResultsRome} and \ref{tab:decisionTreeResultsParis} show that the classifiers demonstrate moderate performance with respect to accuracy. However, from the precision and recall measures we can observe that its performance suffers for different classes. The precision values are undefined because there is atleast one class for which the classifier doesn't declare any instance as a member of that class. Additionally the lower average value of recall gives us an idea that the classifier may perform well for one class whereas it performs poorly for others. The reason behind such result is the imbalance between the classes in our dataset. Class $C_{4.5}$ has 573 instances whereas class $C_{2.5}$ has only 2 instances. Similarly in Paris dataset $C_{4.5}$ and $C_{4.0}$ contains majority of the locations. In subsequent sections we have discussed steps to handle this issue. 

\subsection{Applying oversampling technique to handle overfitting} \label{sec:SMOTEresults}
In order to handle the performance issues due to imbalance in dataset, we have applied state-of-the art oversampling technique namely SMOTE \cite{SMOTE}. SMOTE uses instances of minority class to generate synthetic instances of corresponding class. We have applied SMOTE for each class to generate a balanced dataset with each class having around 580 instances similar to that of class $C_{4.5}$. Similarly we have balanced out the classes in the Paris dataset. Although any other oversampling techniques could have been chosen and they are reported to have competitive performances \cite{AminOversamplingComparison} on various datasets, to avoid complexity we have chosen to apply SMOTE. The oversampled dataset was used to train classifiers and the result of such classifiers can be found in Table \ref{tab:oversampleResultsRome} and \ref{tab:oversampleResultsParis}. 
\begin{center}
	\begin{table}[!h]
		\caption{Classifier accuracy, precision and recall of decision tree classifiers on oversampled Rome dataset with 3417 instances.}
		\label{tab:oversampleResultsRome}
		\centering
		\begin{tabular}{||c | c | c | c||}
			\hline
			Classifier & Accuracy & Precision & Recall\\
			\hline\hline
			J48 &	70.00 &	69.60 &	69.81 \\
			\hline
			REPTree &	67.37 &	65.97 &	67.10 \\
			\hline
			RF &	77.82 &	76.94 &	77.63 \\
			\hline
		\end{tabular}
	\end{table}
\end{center}
\vspace{-2.5em}
\begin{center}
	\begin{table}[!h]
		\caption{Classifier accuracy, precision and recall of decision tree classifiers on oversampled Paris dataset with 1598 instances.}
		\label{tab:oversampleResultsParis}
		\centering
		\begin{tabular}{||c | c | c | c||}
			\hline
			Classifier & Accuracy & Precision & Recall\\
			\hline\hline
			J48 &	62.20 &	66.45 &	67.40 \\
			\hline
			REPTree &	58.07 &	61.86 &	64.61 \\
			\hline
			RF &	69.77 &	72.35 &	74.16 \\
			\hline
		\end{tabular}
	\end{table}
\end{center}
\vspace{-2.5em}
\subsection{Improving classifier accuracy with ensemble method}\label{sec:ensembleResult}
In order to improve classifier performance, we have applied state-of-art ensemble technique on decision tree classifiers. We have used both Bagging and Boosting technique on the decision tree classifier. Bagging is an ensemble learning method that learns classifier on several different distributions of the training set and uses all the classifiers for classification and applies majority voting to get the desired prediction. On the other hand, in each iteration of boosting technique, it tries to learn classifier on the samples that was wrongly classified using previous classifiers and assign a corresponding weight to each classifier. Both bagging and boosting helps to reduce error rate of an individual classifier. The classifier accuracy after the application of ensemble methods are reported in Table \ref{tab:ensembleResultsRome} and \ref{tab:ensembleResultsParis}.
\begin{center}
	\begin{table}[!h]
		\caption{Classifier accuracy, precision and recall of decision tree classifiers with bagging and boosting technique, on oversampled Rome dataset with 3417 instances.}
		\label{tab:ensembleResultsRome}
		\centering
		\begin{tabular}{||c | c | c | c||}
			\hline
			Classifier & Accuracy & Precision & Recall\\
			\hline\hline
			Bagging with J48 &	76.59 &	75.61 &	76.35 \\
			\hline
			Bagging with REPTree &	74.25 &	72.78 &	73.96 \\
			\hline
			Bagging with RF &	79.48 &	78.60 &	79.27 \\
			\hline
			Boosting with J48 &	76.76 &	76.18 &	76.58 \\
			\hline
			Boosting with REPTree &	74.57 &	73.98 &	74.37 \\
			\hline
			Boosting with RF &	79.10 &	78.32 &	78.93 \\
			\hline
		\end{tabular}
	\end{table}
\end{center}
\vspace{-2.5em}
\begin{center}
	\begin{table}[!h]
		\caption{Classifier accuracy, precision and recall of decision tree classifiers with bagging and boosting technique, on oversampled Paris dataset with 1598 instances.}
		\label{tab:ensembleResultsParis}
		\centering
		\begin{tabular}{||c | c | c | c||}
			\hline
			Classifier & Accuracy & Precision & Recall\\
			\hline\hline
			Bagging with J48 &	68.27 &	71.25 &	73.85 \\
			\hline
			Bagging with REPTree &	65.02 &	67.82 &	71.11 \\
			\hline
			Bagging with RF &	72.22 &	74.28 &	77.59 \\
			\hline
			Boosting with J48 &	70.65 &	73.70 &	75.17 \\
			\hline
			Boosting with REPTree &	64.27 &	68.79 &	69.64 \\
			\hline
			Boosting with RF &	73.78 &	75.62 &	78.07 \\
			\hline
		\end{tabular}
	\end{table}
\end{center}
\vspace{-2.5em}
\section{Future works and conclusion}
In this work, we aimed at using machine learning techniques to score each location with an aesthetic score, which will facilitate urban planning, route recommendation, tour planning etc. We focused on Flickr social metadata to predict TripAdvisor ratings. Having accumulated two real world datasets with 850 locations from Rome and 650 locations from Paris we trained decision tree classifiers to predict the aesthetic score of each location. However, it was observed that the ratings received from TripAdvisor have implicit skewness. The imbalance in dataset resulted in performance issues. Removing the issues of imbalanced dataset using SMOTE technique our proposed models achieved as high as 79\% and 73\% accuracy on Rome and Paris dataset respectively. Future works may focus on finding aesthetic routes through a city or assigning aesthetic score to each  route available through various trajectory datasets. Another direction of study can be attempting to predict aesthetic rating of locations by analyzing Flickr photos rather than assessing their metadata.

\balance{}

\bibliographystyle{SIGCHI-Reference-Format}
\bibliography{scenicpath}

\end{document}